\documentclass[aps,pra,superscriptaddress,twocolumn]{revtex4-1}
\usepackage{bm} 
\usepackage{graphicx} 
\usepackage{amsmath}
\usepackage{amsthm,stmaryrd}
\usepackage{amssymb} 
\usepackage{epstopdf} 
\usepackage{txfonts}
\usepackage{color}
\usepackage{xcolor}
\usepackage{subfigure}
\usepackage[T1]{fontenc}

\newcommand{\ket}[1]{|{#1}\rangle}
\newcommand{\bra}[1]{\langle{#1}|}
\newcommand{\mean}[1]{\langle{#1}\rangle}

\begin{document}

\title{Perfect wave-packet splitting and reconstruction in a one-dimensional
lattice}

\author{Leonardo Banchi}
\affiliation{Department of Physics and Astronomy, University College London, Gower Street, WC1E 6BT London, United Kingdom}

\author{Enrico Compagno}
\affiliation{Department of Physics and Astronomy, University College London, Gower Street, WC1E 6BT London, United Kingdom}

\author{Sougato Bose}
\affiliation{Department of Physics and Astronomy, University College London, Gower Street, WC1E 6BT London, United Kingdom}

\date{\today}

\begin{abstract}
  Particle delocalization is a common feature of quantum random walks in 
  arbitrary lattices. However, in the typical scenario a particle spreads over
  multiple sites and its evolution is not directly useful for 
  controlled quantum interferometry, as may be required for technological
  applications. 
  In this paper we devise a strategy to perfectly split the wave-packet of  an
  incoming particle into two components, each propagating in opposite
  directions, which reconstruct the shape of the initial wavefunction after a 
  particular time $t^*$.
  Therefore, a particle in a delta-like initial state becomes 
  exactly delocalized between two distant sites after $t^*$. 
  We find the mathematical conditions to achieve the perfect splitting which are
  satisfied by viable example Hamiltonians with static site-dependent 
  interaction strengths. 
  Our results pave the way for the generation of  peculiar many-body interference
  patterns in a many-site atomic chain (like the Hanbury Brown and Twiss and quantum
  Talbot effects) as well as for
  the distribution of entanglement between remote sites. 
  Thus, as for the case of perfect state transfer, the perfect wave-packet
  splitting can be a new tool for varied applications. 
\end{abstract}

\maketitle

\section{Introduction}
The quest for a quantum computer is boosting  the development and engineering 
of new sophisticated quantum devices that allow us to observe 
the space-time evolution of its constituents. Indeed,
in recent years several experimental groups successfully measured the quantum
dynamical evolution of particles and/or quasi-particle  hopping in a lattice
\cite{preiss_strongly_2014,fukuhara_quantum_2013,fukuhara_microscopic_2013,sansoni_two-particle_2012,ramanathan_experimental_2011,schreiber_observation_2015,richerme_non-local_2014,schachenmayer2013entanglement,trompeter_visual_2006}.
Due to the inherent nature of quantum mechanics, the evolution of an isolated 
quantum system is represented by a wavefunction $\psi(x,t)$ which describes the
probability amplitude of finding a particle in position $x$ at time $t$. 
Quantum interference can give rise to particular structures and patterns in the space-time 
evolution $|\psi(x,t)|^2$ which are known as ``quantum
carpets'' \cite{kaplan_multimode_2000}, quantum revivals \cite{robinett_quantum_2004}, or quantum Talbot effect 
\cite{berry_quantum_2001}, quantum walks \cite{kempe_quantum_2003,childs_universal_2009,childs_exponential_2003},
and quantum self-imaging \cite{longhi_periodic_2010}.  

An interesting case is when the wavefunction undergoes a revival, namely when after a particular time the shape of the initial wave packet  is almost perfectly reconstructed.
%Aside from its fundamental implication, this is particularly important for connecting 
%and linking distant quantum registers
%\cite{cirac_quantum_1997,ritter_elementary_2012},
%when the revival happens in a different position, far
%from the initial one.
Aside from its fundamental implication, revivals occurring into a different
position, far from the initial one, are particularly important 
for connecting 
and linking distant quantum registers \cite{cirac_quantum_1997,ritter_elementary_2012}. 
On the other hand, 
a lattice of static localized particles  represents an alternative 
paradigm for quantum communication where information carriers are
not physically moving particles but rather 
collective excitations whose space extent is reconstructed at a
different position after a certain time. 
In this respect, spin chains represent one of the most viable 
solution and there are
various protocols to exploit their dynamics 
for transferring states and entanglement between remote sites 
\cite{bose_quantum_2007,nikolopoulos_quantum_2013}.  
The coherent excitation transfer, or in general the wavefunction
reconstruction at a certain time, corresponds to the phase alignment of the eigenstates entering into the wave packet and, as such, can happen only when the energy eigenvalues satisfy certain conditions
\cite{aronstein_fractional_1997,kay_perfect_2010}. 
Some models admitting a perfect 
\cite{albanese_mirror_2004,aronstein_fractional_1997,yung_perfect_2005} or almost perfect
\cite{godsil_number-theoretic_2012,banchi_ballistic_2013,banchi_long_2011,yao_robust_2011} reconstruction have been explicitly constructed. On the other hand, if the phase alignment is only between particular subsets of the energy eigenstates, then 
the wavefunction is split into a
superposition of copies of the initial wave-packet, each separated by a certain distance. This effect is known as fractional revival
\cite{aronstein_fractional_1997,chen_fractional_2007,robinett_quantum_2004}, or fractional Talbot effect
\cite{berry_quantum_2001}.

In this paper we engineer a chain with nearest neighbor interactions 
to obtain a perfect wave-packet splitting and reconstruction during a 
ballistic evolution.  
In other terms, if $\psi(x,t{=}0){=}f(x)$ is the shape of the 
initial wave-packet, 
at the revival time $t^*$ the wavefunction is 
$\psi(x,t){=}\frac{f(x{-}vt^*){+} f(x{+}vt^*)}{\sqrt 2}$, where
$v$ is the group velocity defined by the energy eigenvalues.
%Although the particle splitting and reconstruction can be seen as 
%a particular fractional revival, our purpose is to minimize the time between to 
%create the target state.  Indeed, 
While in general the revival time is connected to specific algebraic properties of the
spectrum and might be very long, in our case the splitting happens on a time
which is dictated by the group velocity of excitations and, as such, scales only
linearly with the distance. Our method is therefore specifically targeted for
applications where a smaller operational time is particularly important for
neglecting the interaction with the surrounding environment. 
Recently, it has been shown that the wavefunction of a one-dimensional excitation 
can be split into a transmitted and reflected components by introducing
localized impurities  \cite{compagno_beam_2014,fogarty_effect_2013,gertjerenken_bright-soliton_2013}, or via suitably designed time-dependent control fields \cite{makin_spin_2012}. 
Here we focus on a different strategy aiming at obtaining a {\it perfect} fractional revival. 

The generalization of the fractional revival to a many-particle setting has many important
applications. As far as identical particles (bosons/fermions) are concerned, it allows one to define a perfect effective beam splitter operation between distant sites and then to observe multi-particle Hanbury Brown and Twiss interference effects \cite{brown_correlation_1956}, such as perfect bunching or anti-bunching. 
%Moreover, it 
%may pave the way for the lattice realization 
%of the linear optical quantum computer \cite{knill_scheme_2001}.
As for spin systems, that the perfect fractional revival can be
used to generate dynamically long-distance entanglement,
a topical application which may be tested experimentally with current technology
\cite{mitra_spin_2015,sahling_experimental_2015}.
Indeed, the use of particle delocalization to generate entanglement is 
particularly evident in a single excitation setup, 
namely when there is a single spin in 
the state $\ket\uparrow$ while
all the other spins are in the state $\ket\downarrow$. If the wave-packet of
this excitation is perfectly split and reconstructed in two distant sites $n$
and $m$, then the final state of the spins pair $(n,m)$ is 
$(\ket{\uparrow\downarrow}_{nm}+\ket{\downarrow\uparrow}_{nm})/\sqrt 2$, namely
a maximally entangled Bell state. We show that this reasoning 
can also be used in a multiple excitation scenario to dynamically 
generate a maximal set of Bell pairs in a spin
chain setup, and to provide a more general version of previous proposals
\cite{alkurtass_optimal_2014,di_franco_nested_2008}.

This paper is organized as follows. %In section II we study a homogeneous one 
%dimensional system to discuss analytically the splitting of a wave-packet by
%a single impurity in the central site. 
In section II, we define the mathematical conditions which allows a particular
fractional revival, namely the perfect splitting and reconstruction of an
incoming wave packet, and we propose a numerical algorithm 
to find suitable Hamiltonians which fulfill these conditions. 
Interesting applications are then analyzed in section
III in a many-particle setting. 
In particular, we discuss bunching/anti-bunching effects in atoms trapped 
in an optical lattice and the dynamical generation of entanglement in 
spin chains interacting with nearest-neighbor XY couplings. 
Conclusions are drawn in section IV. 

\section{Perfect splitting with engineered couplings}
We study a one-dimensional quantum walk in a lattice with nearest-neighbor 
engineered interactions described by the Hamiltonian 
\begin{align}
  H_p=-\left(\sum_{n=1}^{L-1} J_n\ket n \bra{n{+}1} + {\rm h.c.}\right) -
  \sum_{n=1}^L B_n \ket n \bra n~,
  \label{e.hameng}
\end{align}
where $\ket n$ represents the state where a particle is in the $n$-th site, and 
$L$ is the length of the chain. 
To find the mathematical conditions for a perfect splitting and reconstruction, 
we first focus on the requirements to achieve perfect state transfer.
To perfectly transfer an excitation from site 1 to site $L$ the coefficients 
$J_n$ and $B_n$ have to satisfy some conditions (see e.g. Ref.\cite{kay_perfect_2010}).  
%Perfect state transfer requires some specific conditions on the 
%eigenvalues of $H_p$. 
Firstly, the Hamiltonian has to be {\it mirror symmetric},
i.e. $J_{L-n}{=}J_n$ and $B_{L+1-n}{=}B_n$ for any $1{\leq} n {\leq} L$.
The mirror symmetry imposes some relations between the
eigenvectors of the Hamiltonian
\cite{yung_perfect_2005}: if the eigenvalues $E_k$ of $H_p$ 
are ordered such that $E_k {<} E_{k+1}$, then
\begin{align}
  O_{Lk}=(-1)^k O_{1k}~,
  \label{e.mirrorsym}
\end{align}
where $H_p=OEO^T$ is the eigenvalue decomposition of $H_p$. 
The second requirement is that the energy 
eigenvalues $E_k$ satisfy the relation
\begin{align}
  e^{-i E_k t^*} = (-1)^k e^{i\alpha}~,
  \label{e.perfc}
\end{align}
where $t^*$ is the transmission time and $\alpha$ is an arbitrary phase.
 Here we consider $\alpha{=}0$, namely ${\rm Tr} H{=}0$. 
Among the analytic solutions 
of \eqref{e.perfc}, the simplest one is given by the coupling pattern 
\cite{christandl_perfect_2004,albanese_mirror_2004} 
\begin{align}
  J_n^{\rm PST}&=\frac{\pi J}{2L}\sqrt{n(L-n)}~,& B_n&=0~,
  \label{e.christandl}
\end{align}
which implements perfect state transfer (PST) at $t^*{=}L{/}J$.
Other solutions can be obtained numerically using inverse eigenvalue 
algorithms \cite{kay_perfect_2006,kay_perfect_2010,bruderer_exploiting_2012}.
If the eigenvalues and eigenvectors of $H_p$ satisfy Eqs.\eqref{e.mirrorsym}
and \eqref{e.perfc}, then 
\begin{align}
  \bra n e^{-iH_pt^*} \ket m &= \sum_k O_{n,k}O_{m,k} e^{-iE_k t^*}
   =\sum_k O_{n,k}O_{m,k} (-1)^k
   \nonumber\\&=\sum_k O_{n,k}O_{L+1-m,k} =\delta_{n,L+1-m}~,
  \label{e.perftrans}
\end{align}
namely an excitation initially located in site $m$ is perfectly transferred to site $L{-}m{+}1$ after a time $t^{*}$.

In a similar fashion, a perfect wave-packet splitting and reconstruction 
can be obtained when the eigenvalues of $H_p$ satisfy
\begin{align}
  e^{-i E_k t^*} = \cos\theta + i(-1)^k \sin\theta~,
  \label{e.perfsplitter}
\end{align}
for some angle $\theta$. 
Indeed, by repeating the calculation \eqref{e.perftrans} one finds
$  \bra n e^{-iH_pt^*} \ket m {=} \cos\theta \;\delta_{nm} {+} i \sin\theta\;\delta_{n,L+1-m}~,$ namely 
\begin{align}
  \ket m \xrightarrow{t^*} \cos\theta \ket m+i\sin\theta\ket{L{-}m{+}1}~.
  \label{e.transf}
\end{align}
%In \cite{compagno_beam_2014} the authors have shown that an almost perfect 
%splitting can be achieved by using the coupling pattern \eqref{e.christandl}
%with a localized impurity $B\ket N\bra N$ as in the previous section. 
The eigenvalue relations \eqref{e.perfsplitter} are one of the main result 
of this paper. By properly choosing $\theta$ it is possible to balance the
reconstruction on distant sites, as show in Eq.\eqref{e.transf}, and for
$\pi{=}\pi{/}4$ one obtains the perfect delocalization between distant sites 
of an initially localized wave packet. 
The coupling pattern to satisfy Eq.\eqref{e.perfsplitter}
can be obtained using inverse eigenvalue techniques.
From the conceptual point of view an inverse eigenvalue problem deals 
with finding the zeros of the highly non-linear function 
$f(\lambda){=}E(\lambda){-}\tilde E$, 
where the vector $E(\lambda)$ contains the ordered eigenvalues of the
Hamiltonian $H_p(\lambda)$ with parameters 
$\lambda$, and the vector $\tilde E$ contains the target spectrum.  
Among the algorithms to find the optimal parameters
\cite{parlett_symmetric_1998,chu_inverse_1998}, 
the most used one relies on the application of the 
Newton method to find the zeros of $f(\lambda)$.
The Newton method starts with an initial guess $\lambda^{(0)}$ and updates it according to the rule \cite{friedland_formulation_1987}
\begin{align}
  \mathcal J(\lambda_n) [\lambda^{(n+1)}-\lambda^{(n)}] = f(\lambda^{(n)})~,
  \label{e.newton}
\end{align}
where the matrix, with elements
\begin{align}
  \mathcal J_{mk}(\lambda^{(n)}) =\frac{\partial f_m(\lambda^{(n)})}{
  \partial \lambda^{(n)}_k} 
  = \bra m O(\lambda^{(n)})^T\frac{\partial H_p(\lambda^{(n)})}{\partial
    \lambda^{(n)}_k} O(\lambda^{(n)})\ket k~,
  \label{e.jacobian}
\end{align}
is the Jacobian matrix and $H_p(\lambda){=}O(\lambda)E(\lambda)O(\lambda)^T$ is the eigenvalue decomposition of $H_p(\lambda)$. 
The linear system \eqref{e.newton} has a unique solution provided that $\mathcal
J$ is an invertible matrix. This in turn implies that the number of parameters
have to match the number of eigenvalues, i.e. the dimension of the matrix. 

The mirror symmetric Hamiltonian \eqref{e.hameng} has $L$ independent
parameters, being $L$ the number of sites. Indeed, because of the mirror
symmetry, when $L{=}2N$ (being $N$ an integer) there are $N$ independent values of $J_n$ and $N$ independent values of $B_n$. On the other hand, when $L{=}2N{+}1$, there are $N$
independent values of $J_n$ and $N{+}1$ independent values of $B_n$. We apply
inverse eigenvalue techniques to find the coupling pattern which allows a
perfect balanced splitting of the wave-packet. The latter is 
obtained by imposing the condition \eqref{e.perfsplitter} with $\theta{=}\pi{/}4$,
so the target eigenvalues are 
\begin{align}
  \frac{L\tilde E}J= \left(\dots, -\frac\pi4,
   \frac\pi4,
   -\frac\pi4+2\pi,
   \frac\pi4+2\pi,
   -\frac\pi4+4\pi,
   \dots\right)^T
  \label{e.target}
\end{align}
where, without loss of generality, we have imposed $t^*=L/J$.
Because $f(\lambda)$ is in general a non-convex function, possibly with many
local minima, inverse eigenvalue problems
are known to converge only if the initial guess $\lambda^{(0)}$ is not too far
from the ideal set of parameters $\tilde\lambda$ for which $E(\tilde\lambda)
{=}\tilde E$ \cite{friedland_formulation_1987}.
%Motivated by the solution found in the previous section, where the splitting 
%was implemented by a perturbation, w
We guess that the optimal parameters for a perfect wave packet splitting
are given by a local perturbation of the fully engineered chain 
which guarantees perfect state transfer, so we use the 
coupling pattern \eqref{e.christandl} as an initial condition. 

The algorithm quickly converges to an optimal parameter set and hereinafter  we 
called $J_n^{\rm split}$ and $B_n^{\rm split}$ the obtained optimal couplings and local
fields. Surprisingly,
we find that for even $L$ the algorithm always converges to a solution where
$B_n^{\rm split}{=}0$, while for odd $L$ the local fields $B_n^{\rm split}$ 
are different from zeros
especially near the center of the chain. 
For example, 
the Hamiltonians for $L{=}5{,}6$ are shown in Appendix A. 
\begin{figure}[t]
 \centering
\subfigure[\ Even length chain]{\includegraphics[width=.45\textwidth]{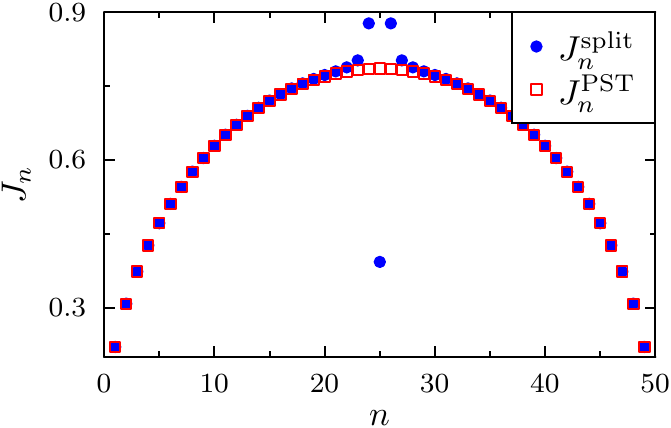}\label{fig:Perf50}}
\subfigure[\ Odd length chain]{\includegraphics[width=.45\textwidth]{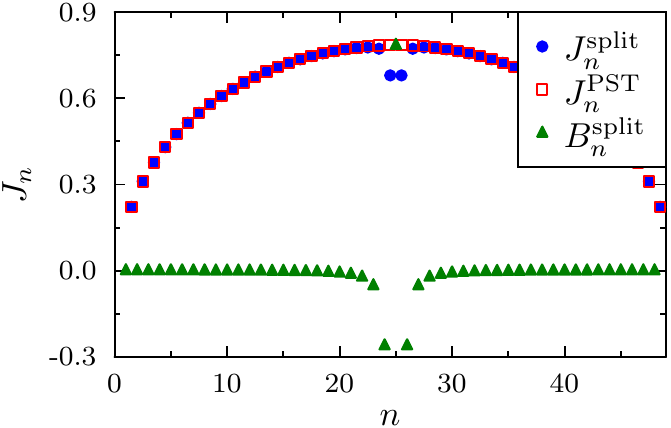}\label{fig:Perf49}}
\caption{Comparison between the perfect wave-packet splitting couplings
  $J_n^{\rm split}$ and the perfect state transfer couplings $J_n^{PST}$ in
  Eq.\eqref{e.christandl} for (a) an even chain and for an (b) odd chain. Only
  in  the latter case the perfect splitting requires also the engineering of 
  a field profile $B_{n}^{split}$.}
\end{figure}
The output of the algorithm is shown in Fig.\ref{fig:Perf50} for $L{=}50$, and in Fig.\ref{fig:Perf49} for $L{=}49$. As it is clear, both for $L$ even and odd the coupling patterns $J_{n}^{\rm split}$ for  perfect wave-packet splitting are similar to the coupling pattern $J_{n}^{PST}$, in formula \eqref{e.christandl}, for perfect state transfer:
the only difference being the presence of few impurities at the center of the chain. 
Moreover, %while for even $L$ the algorithm always outputs $B^{\rm split}_n=0$, 
for odd $L$ one requires also the engineering of the local fields
$B_n^{\rm split}$ according to some particular profile. The resulting field pattern is
constant far from the center of the chain and has a particular oscillatory
profile near the central sites. 

\section{Applications}
\subsection{Perfect bunching/anti-bunching in a bosonic lattice}
As a concrete application of the results of the last section we consider a model of hopping particle in a one-dimensional lattice, described by a Bose-Hubbard Hamiltonian with site dependent parameters
\begin{equation}
  \mathcal H=-\sum\limits_{n=1}^{L} J_n \left( a_{n}^\dag 
  a_{n+1}+ \text{h.c.}\right)+ \sum\limits_{n=1}^{L}U_n  n_n\left(  n_n
  -1\right) - \sum_{n=1}^{L} B_n  n_n .
\label{eq:BHHamiltonian}
\end{equation}
Here $J_n{=}J_n^{\rm split}$ are the tunneling matrix elements, $U_n$ is the
onsite interaction and $B_n{=}B_n^{\rm split}$ is the chemical potential, $a_n$ is
the boson annihilation operator and $n_j{=}a_j^\dagger a_j$. The Hamiltonian 
\eqref{eq:BHHamiltonian} accurately describes cold bosonic atoms in optical lattices \cite{jaksch_cold_1998,greiner_quantum_2002}, and it also models fermions \cite{jordens_mott_2008,modugno_production_2003} and hard-core bosons \cite{rigol_universal_2004} dependently on the onsite interaction values.
%Alternative implementations are systems of interacting polaritons in coupled arrays of cavities \cite{hartmann_strongly_2006} and Josephson junctions \cite{van_der_zant_field-induced_1992}
%}.  In the optical lattice implementation t
The tuning of the site dependent coupling constants in \eqref{eq:BHHamiltonian} is achieved via addressable optical lattices \cite{wang_fault-tolerant_2013}, created projecting an electric field profile via holographic masks \cite{bakr_quantum_2009,boyer_dynamic_2006} or via micro-mirror devices \cite{preiss_strongly_2014}. Initialization and read-out of single atoms are achieved exploiting single-particle addressing techniques \cite{preiss_strongly_2014,weitenberg_single-spin_2011,endres_single-site-_2013,gross_microscopy_2014,schreiber_observation_2015} while magnetically induced Feshbach resonances allow a global control of the onsite interaction acting on the collisional coupling constants values 
\cite{sanders_bound_2011}. For instance the non-interacting regime $U_n{=}0$ has
been recently achieved with this technique using $\rm Cs$ atoms loaded in a one-dimensional optical lattice \cite{meinert_interaction-induced_2014}. 

Thanks to the techniques developed in this paper, the coupling profile produces
a splitting of a single particle wavefunction, which is reconstructed at the
transfer time $t^*$ as two copies of the initial wavepacket with probability
$1{/}2$ each. %This effect known as fractional revivals \cite{robinett_quantum_2004}, allows us to realize multi-particle quantum interference effects among several single-particle wavepackets.
More precisely when the coupling pattern $J_n^{\rm split}$ is implemented, the
wavefunction of a bosonic atom initially onsite $n$ is split by the impurity
pattern at the center of the lattice and, at the transfer time $t^*$, that
particle is perfectly delocalized between two mirror symmetric sites, $n$ and
$L{-}n{+}1$. If an another particle was in the lattice in position $m$, after $t^*$
it would be delocalized between the sites $m$ and $L{-}m{+}1$. When two
particles are initially in two mirror symmetric sites, i.e. $m{=}L{-}n{+}1$ the
dynamics generates multi-particle Hanbury Brown and Twiss correlations 
\cite{brown_correlation_1956}
at $t^*$. Indeed, in the free boson case, namely when $U_n {=} 0$, because of the symmetries of the bosonic wave-function, after a time $t^*$ the state becomes
\begin{align}
 \ket{\psi(t^*)}= \frac{\ket{2}_n\ket0_m+\ket0_n\ket2_m}{\sqrt 2}
 \equiv\ket{\psi_b}~,
  \label{e.hongoumandel}
\end{align}
i.e. the output state consists of a superposition of two bosons being in site $n$ and two bosons being in site $m{=}L{-}n{+}1$. 
This ``bunching'' effect is the celebrated Hong-Ou-Mandel effect (HOM) \cite{hong_measurement_1987} which has been observed recently exploiting the coherent evolution of two particles in a single double-well tunneling model \cite{kaufman_two-particle_2014}. 
With the results presented in this paper, because of the perfect reconstruction of wave-packets at the transfer time, it is possible to achieve a perfect bunching between arbitrary distant sites of an optical lattice. On the other hand, in the strong interacting case, namely in the hard-core boson
limit $U_n{=}\infty$, the final state is $\ket{\psi(t^*)}{=}\ket1_m\ket1_n$, i.e. there is one particle in position $n$ and one particle in position $m$. %\textcolor{red}{In Fig. \ref{fig:PerfectEven}  we show the joint probability $P_{jk}(t)$ to have one boson in the site $j$ and one in $k$, assuming that the two particles were initially sites $1$ and $L$. The dynamics is evaluated by projecting the Hilbert space in the two particle sector as shown in Appendix B. As it can be seen clearly, the Hong-Ou-Mandel effect happens at time $(2k+1)t^*$ for any integer $k$, while at time $2kt^*$ the chain goes back into the initial state. In the fermion case the analogous anti-bunching effect is shown in Fig.\ref{fig:Fermion}.  }
%\begin{figure}[t]
%\includegraphics[width= 8 cm]{PerfectEven}
%\caption{Joint probability $P_{jk}(t)$ for two bosons as function of the transfer time $t^*$ for $L{=}20$ and $U_n{=}0$. }
%\label{fig:PerfectEven}
%\end{figure}
%\begin{figure}[t]
%\includegraphics[width= 8 cm]{PlotPerfectFermion}
%\caption{Joint probability $P_{jk}(t)$ for two fermions/hard-core bosons as function of the transfer time $t^*$ for $L{=}20$ and $U_n{=}0$. }
%\label{fig:Fermion}
%\end{figure}

\subsubsection{Effect of imperfections in tuning the parameters}
In real systems random noise effect, due to environmental variables, and
engineering imperfections in the coupling configurations produce deviations from the
ideal coupling values \cite{wang_fault-tolerant_2013}. The effect of the
coupling randomness, even for non-interacting systems, is to produce a
localization of the eigenstates of the system and consequently to inhibit the
state transfer \cite{keating_localization_2007}. We also mention that recently
it has been shown \cite{stasinska_bose-hubbard_2014,ospelkaus_localization_2006}
that the interaction of bosonic atoms with static fermionic impurities, randomly
distributed in the lattice, may yield a Bose-Hubbard model where the parameters
$J_n$ and $B_n$ are subject to noise. Given the above, we investigate what degree of imperfections  is tolerable in our scheme or, in other terms, what is the precision required in tuning the coupling strengths according to the desired pattern.

\begin{figure}[t!]
\subfigure[Off-Diagonal Noise]{\includegraphics[width= 7 cm]{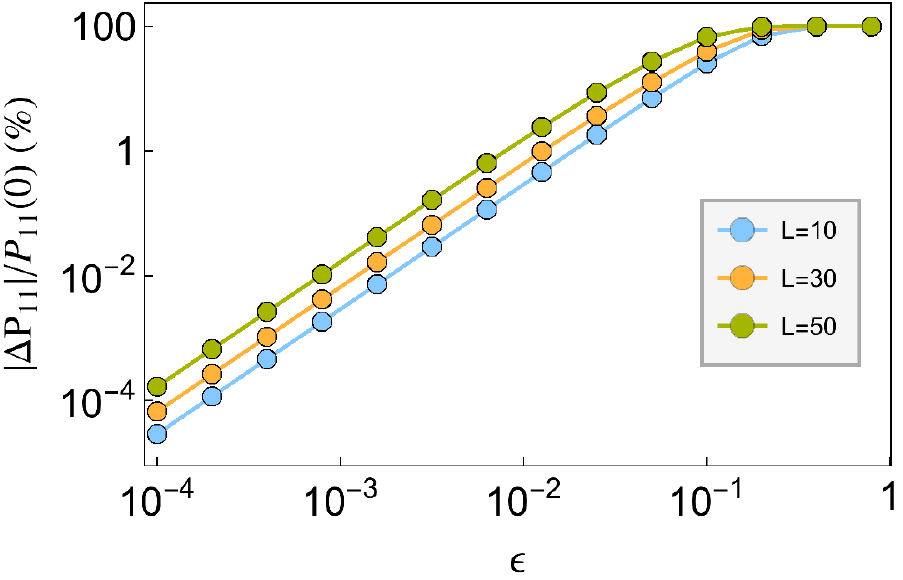}\label{fig:dataU0FuncEps}}
\subfigure[Diagonal Noise]{\includegraphics[width= 7 cm]{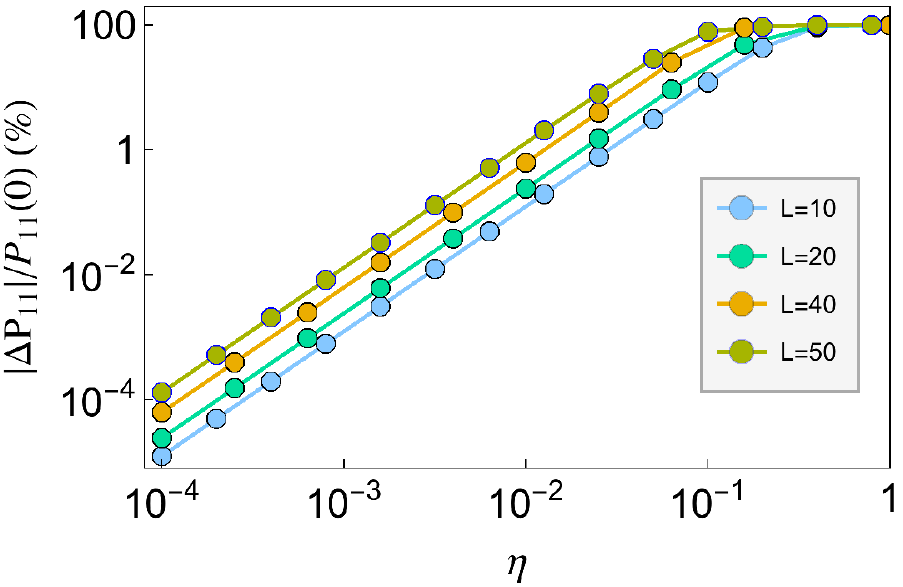}\label{fig:dataU0DiagonalFuncEps}}
\caption{Relative variation of the bunching probability $P_{11}(t{=}t^{*})$ in
the non-interacting regime $U_n{=}0$, in presence of (a) random hopping coupling
strength $\epsilon$ and (b) random diagonal coupling strength $\eta$. Several
chain lengths $L$ are considered.}
\end{figure}

We  firstly include an off-diagonal disorder term (hopping disorder) in the Hamiltonian \eqref{eq:BHHamiltonian} as  $J_{n}{=}J(J_{n}^{split}{+}x_n)$, where $x_n{\in}\left[-\epsilon,\epsilon \right]$ is a uniform random distribution and $\epsilon$ is the perturbation strength \cite{zwick_robustness_2011}. In Fig. \ref{fig:dataU0FuncEps} the relative variation $\vert \Delta P_{11}\vert {/}P_{11}(\epsilon{=}0)$ is shown as function of the degree of disorder $\epsilon$. Here 
$P_{11} {=} |\langle{\psi_b}|{\psi(t^*)}\rangle|^2$, where $\ket{\psi_b}$ is defined
in \eqref{e.hongoumandel}, and 
$\Delta P{\equiv}\vert P_{11}(\epsilon){-}P_{11}(\epsilon{=}0)\vert$ represents the deviation of the bunching probability respect to the ideal case. 
We also consider the effect of diagonal noise $B_j{=}B{+}J x_j$ with $x_n{\in}\left[-\eta,\eta \right]$ in an even site chain. The effect of signal noise is shown in figure Fig. \ref{fig:dataU0DiagonalFuncEps} as function of the noise coupling strength $\epsilon$. 
As clear from Fig. \ref{fig:dataU0FuncEps} and \ref{fig:dataU0DiagonalFuncEps} a power law behavior, under a certain threshold value of $\epsilon$ and $\eta$, characterizes both the deviations due to hopping disorder and due to the diagonal disorder. Clearly, an high degree of disorder produces state localization, which completely destroys the effect.

\subsection{Quantum many-particle carpets}

\begin{figure}[t!]
\includegraphics[height=7 cm]{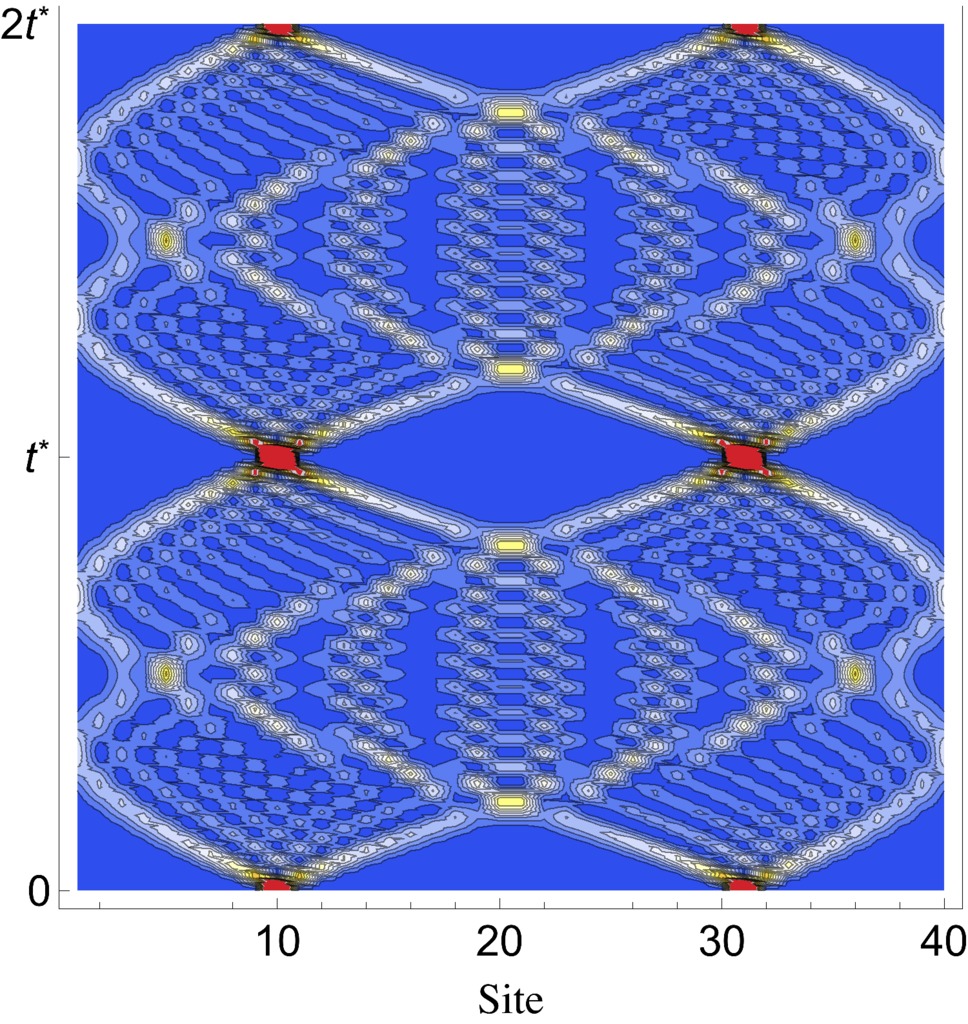}
\hspace{3mm} \includegraphics[height=7.3 cm]{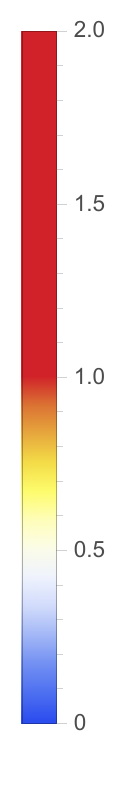}
\caption{Quantum Carpet: space-time evolution of $\langle n_{j}^{2} (t)\rangle $
for a $L{=}40$ one-dimensional chain initialized in
$\ket{\psi_0}{=}a^{\dagger}_{10}a^{\dagger}_{31}\ket{0}$. We consider the
free-boson regime $U_n{=}0$. The white zones are out of the range $[0,0.4]$
which has been chosen for convenience. 
}
\label{fig:1DCarpetsn2}
\end{figure}

\begin{figure*}[t!]
\subfigure[$U{=}0$]{\includegraphics[width=5.8 cm]{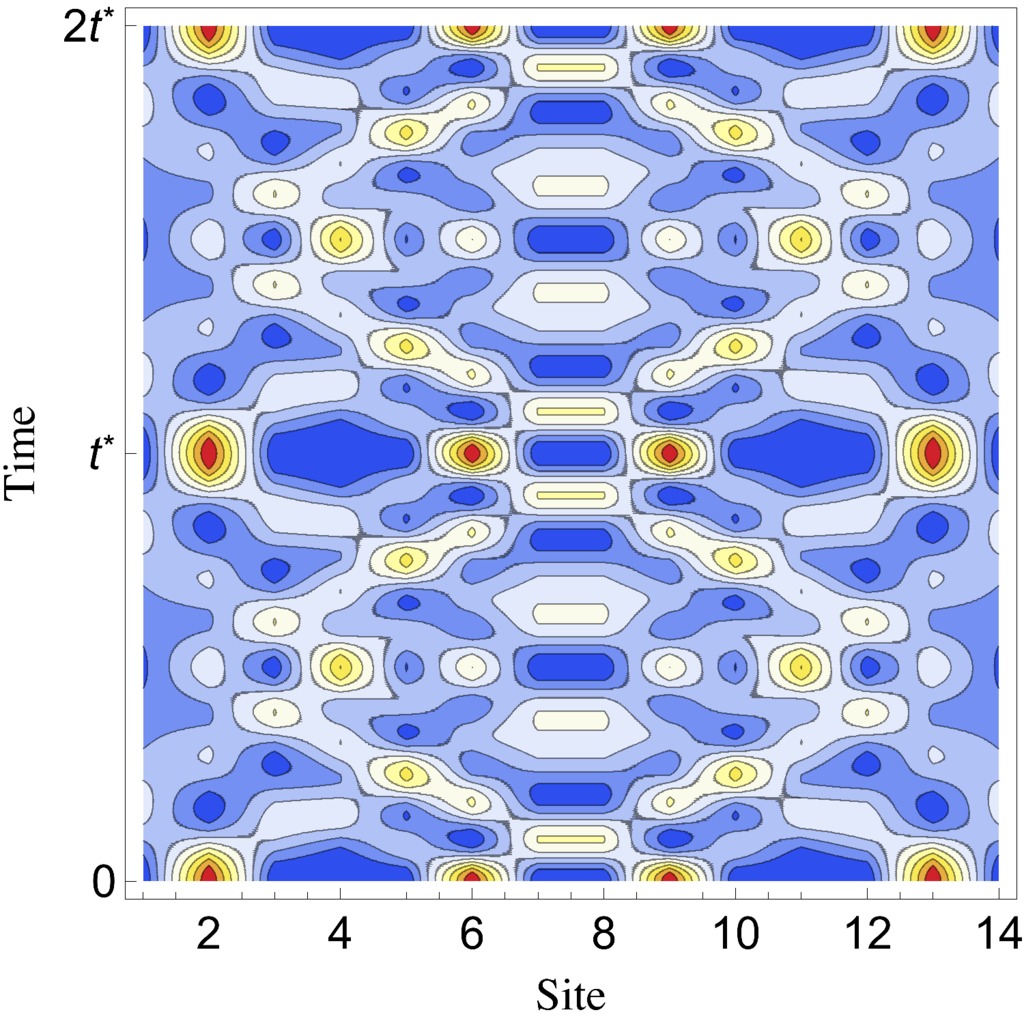}}
\subfigure[$U{=}1$]{\includegraphics[width=5.5 cm]{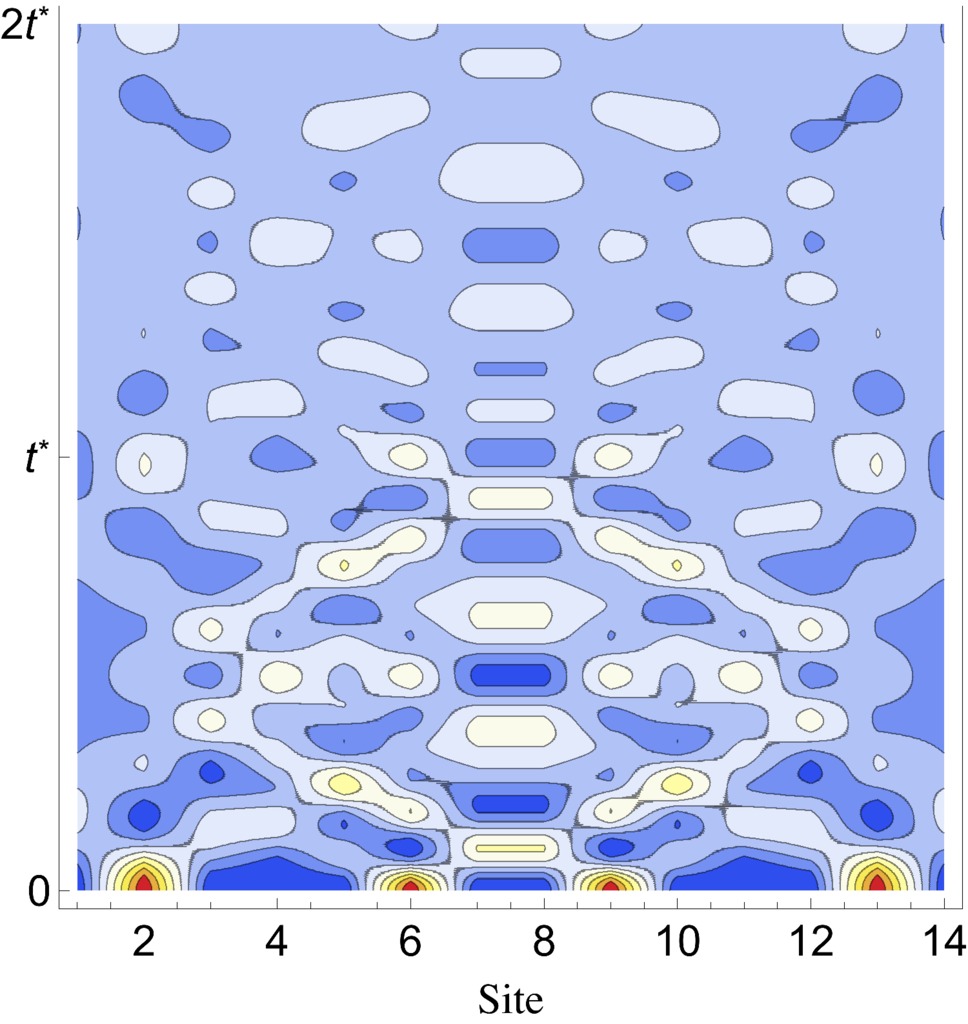}}
\subfigure[$U{=}30$]{\includegraphics[width=5.5 cm]{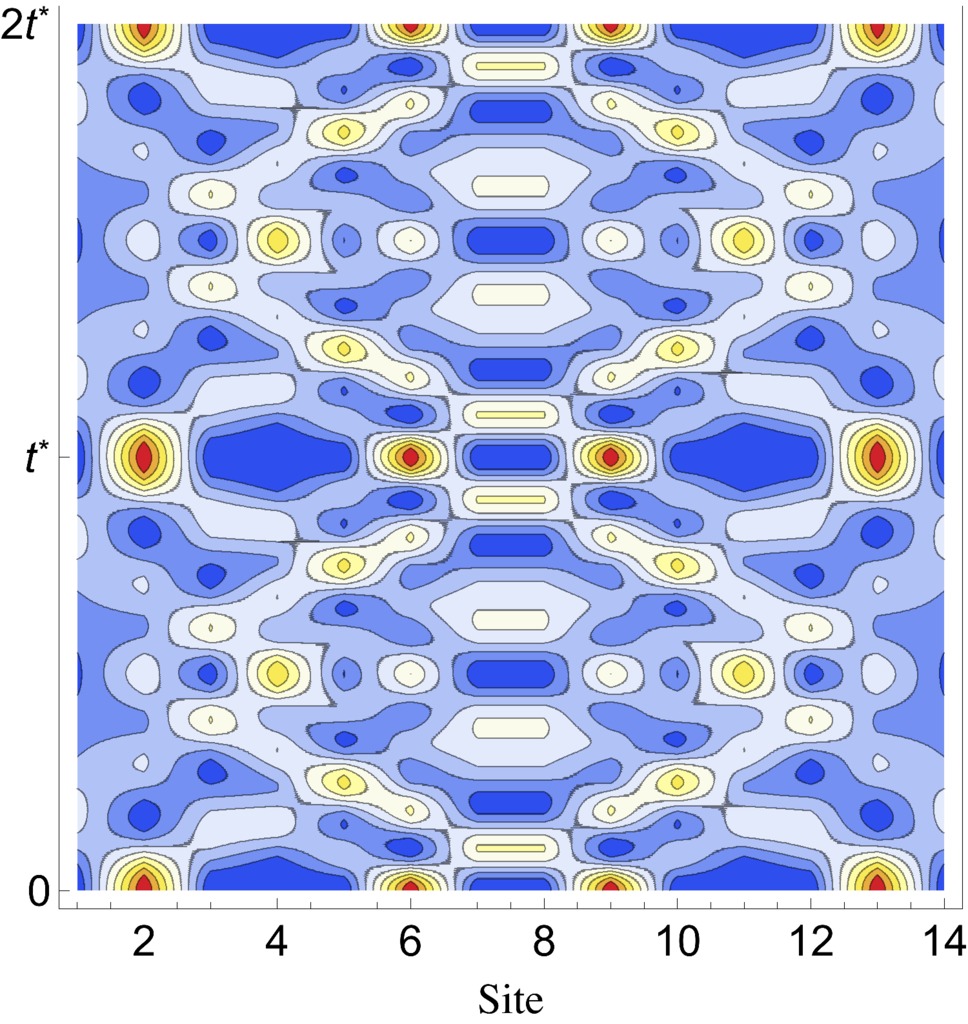} }
\subfigure[$U{=}0$]{\includegraphics[width=5.8 cm]{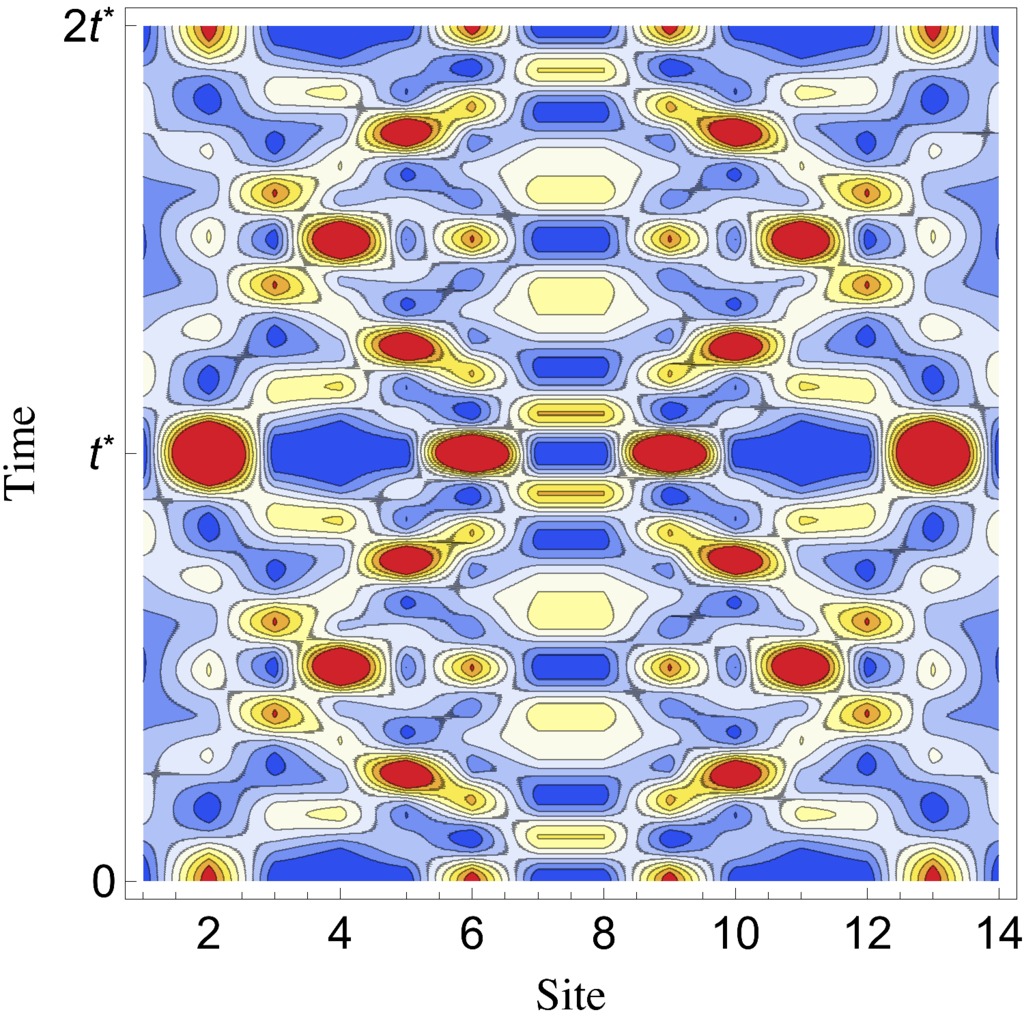}}
\subfigure[$U{=}1$]{\includegraphics[width=5.5 cm]{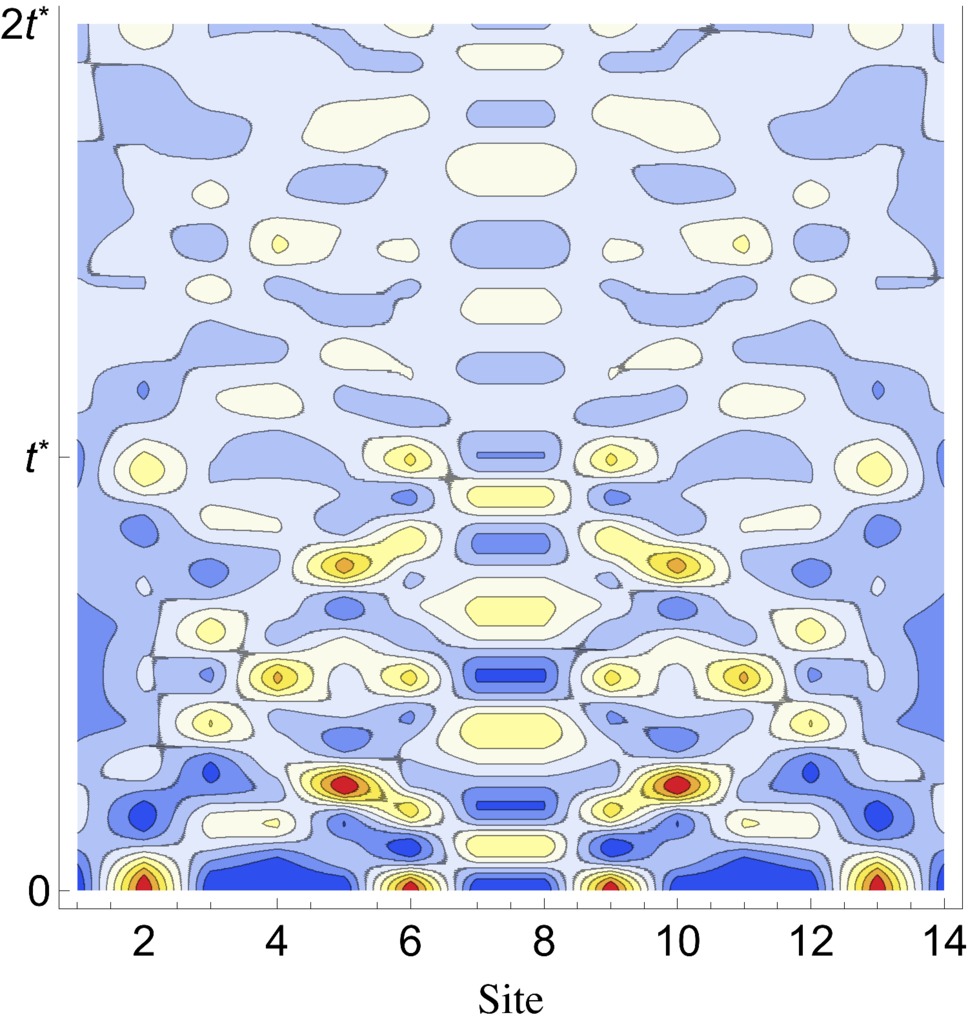}}
\subfigure[$U{=}30$]{\includegraphics[width=5.5 cm]{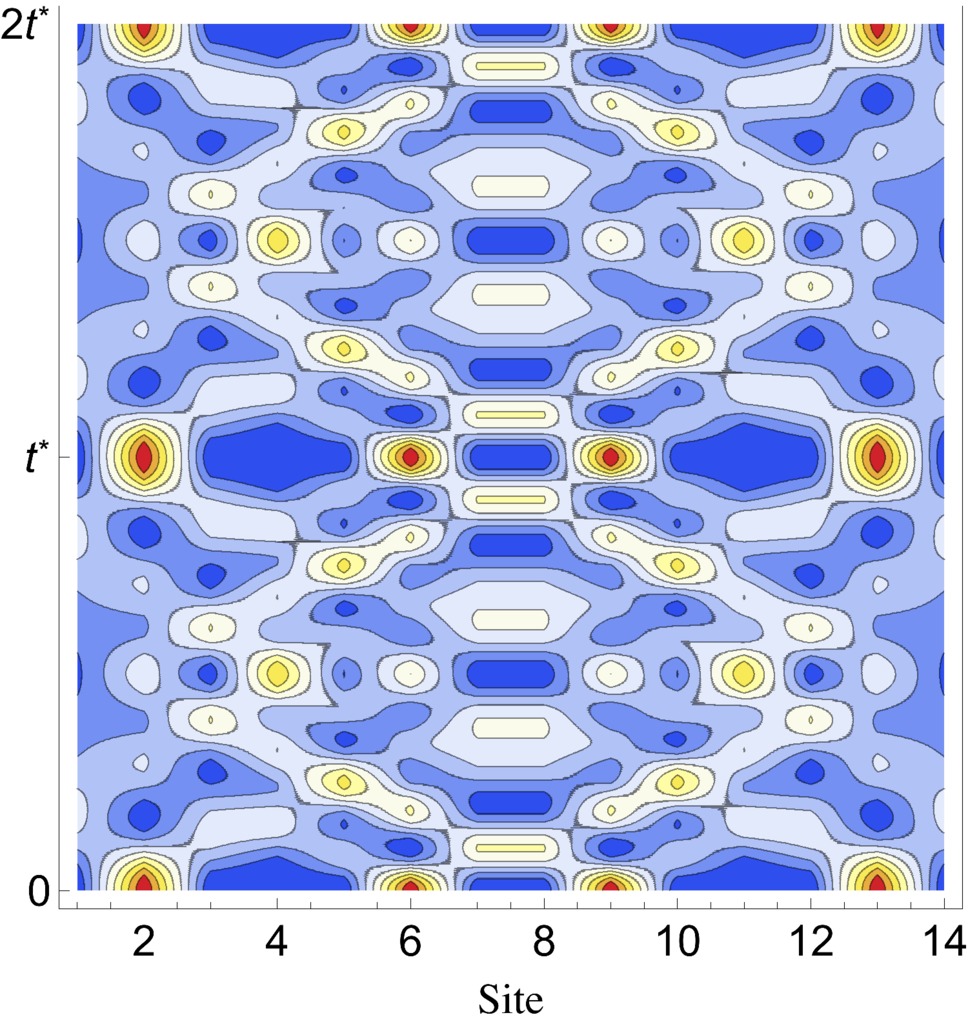} }
\hspace{1 cm}
\subfigure{\includegraphics[width=10 cm]{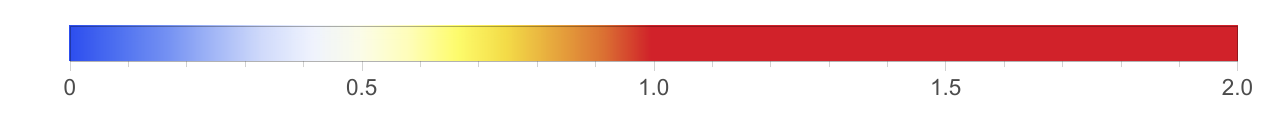} }
\caption{Quantum Carpet due to four particle interference. The initial state 
  $\ket{\psi_0}{=}a_{2}^{\dagger}a_{6}^{\dagger}a_{9}^{\dagger}a_{13}^{\dagger}\ket{0}
  $ contains four bosonic particles. 
  We consider the space-time evolution of $\langle n_{j}(t)\rangle$ in
  figure (a,b,c), and the space-time evolution of $\langle n_{j}^{2}(t)\rangle$
  in figures (d,e,f). The chain length is $L{=}14$ and several value of the
  onsite interaction $U_{n}{=}U$ are considered. Here $t^*$ is the fractional
revival time, while $2t^*$ is the full revival time. The difference between the
first and second row is due to bunching/anti-bunching effects. 
Note the transition from bosonic ($U{=}0$) to fermionic and hard-core boson
($U{=}\infty$)  behavior as a function of $U$.
}
\label{fig:1DCarpetsn4}
\end{figure*}

The perfect reconstruction scheme developed in the previous sections allows generating periodic space-time quantum interference patterns of multi-particles systems known as ``quantum carpet''. 
By using the engineered chain with $J_n^{\rm split}$ and $B_n^{\rm split}$ one
in fact expects a regular temporal pattern in the evolution: the
wave-packets composing the initial state are split into two copies,
reconstructed into different
positions after the time $t^*$, and then they go back to the initial position after a time
$2t^*$. On the other hand, during intermediate times, quantum interference
leads to different behaviors which are expected to be susceptible to the
particle statistics. 
To show this effect, we study the quantum carpet generated by the space-time 
evolution of the mean occupation number $\mean{ n_j(t)}$, or by 
the square occupation mean $\mean{ n_{j}^{2}(t)}$. 
The regular interference pattern of a two particle system is depicted in Fig. \ref{fig:1DCarpetsn2} where we show the expectation value $\langle n_j^2 (t)\rangle$ for two non interacting bosons initially in $\ket{\psi_0}{=}a_{10}^{\dagger}a_{31}^{\dagger}\ket{0} $ in a one-dimensional chain with $L{=}40$.

\begin{figure}[h]
\subfigure[Mean occupation number]{\includegraphics[width= 7 cm]{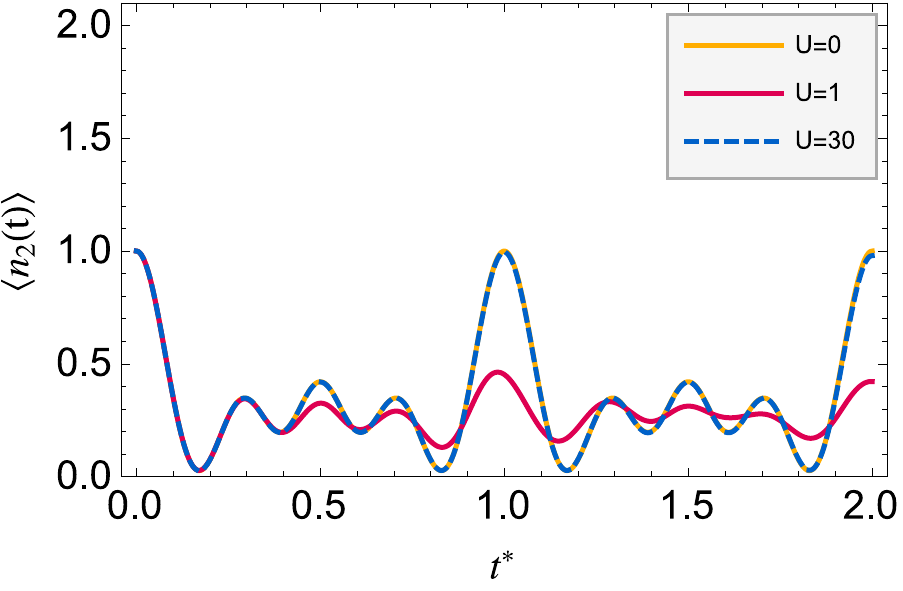}\label{fig:NSquare10}}
\subfigure[Quadratic mean occupation number]{\includegraphics[width= 7 cm]{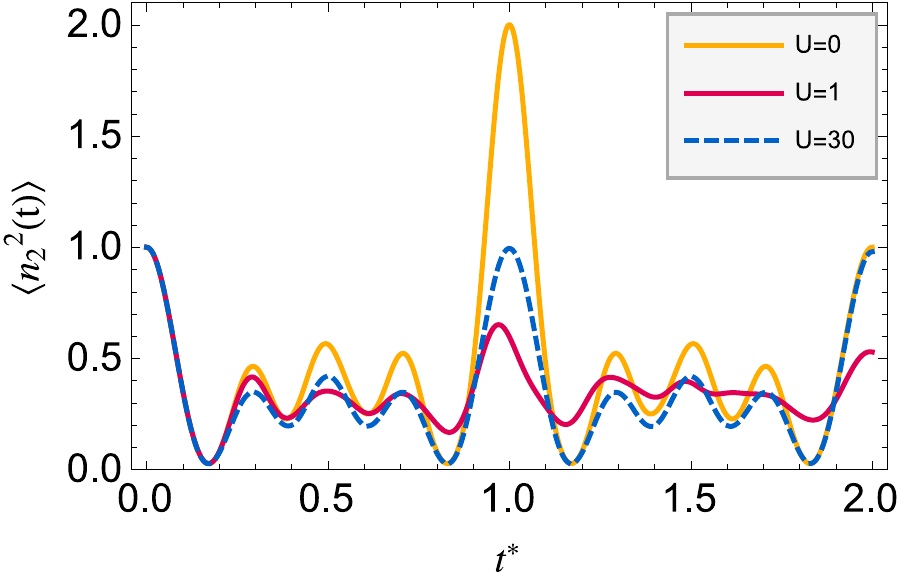}\label{fig:NSquare10}}
\caption{Plot of the (a) mean occupation number and of the (b) quadratic mean occupation number of site $2$ as function of time for several values of the onsite interaction. Here $L{=}14$ and the initial state is $\ket{\psi(0)}{=}a_{2}^{\dagger}a_{6}^{\dagger}a_{9}^{\dagger}a_{13}^{\dagger}\ket{0}$.
Note the transition from bosonic ($U{=}0$) to fermionic and hard-core boson
($U{=}\infty$)  behavior as a function of $U$.
}
%\caption{$\langle n(t)\rangle$ for site $2$ as function of time for a $L{=}14$ chain initialized in $\ket{\psi}{=}a_{2}^{\dagger}a_{6}^{\dagger}a_{9}^{\dagger}a_{13}^{\dagger}\ket{0} $ for several value of the onsite interaction $U$. }
%\caption{$\langle n^2(t)\rangle$ for site $2$ as function of time for a $L{=}14$ chain initialized in $\ket{\psi}{=}a_{2}^{\dagger}a_{6}^{\dagger}a_{9}^{\dagger}a_{13}^{\dagger}\ket{0} $ for several value of the onsite interaction $U$. }
\label{fig:Slice2}
\end{figure}

To highlight more in detail the multi-particle statistical interference effect
we consider a system of four particles, initially in
$\ket{\psi_0}{=}a^{\dagger}_{2}a^{\dagger}_{6}a^{\dagger}_{9}a^{\dagger}_{13}\ket{0}$,
where $L{=}14$. We show in Fig. \ref{fig:1DCarpetsn4}(a) and
\ref{fig:1DCarpetsn4}(d) respectively the
mean occupation number and the quadratic mean occupation number for the non
interacting case and for the strong interacting case in Fig.
\ref{fig:1DCarpetsn4}(c) and \ref{fig:1DCarpetsn4}(f). In the boson case bunching effects are observable at $t^*$ while in both cases a perfect reconstruction of the initial wavepacket happens at $2t^*$. This is evident more clearly in Fig. \ref{fig:Slice2} where we represent the mean occupation number and the quadratic mean occupation number of site $2$ as function of time. 
%The mean particle number in each site $\langle n_j(t)\rangle{=}\langle \psi(t)\vert a_{j}^{\dagger} a_{j}\vert \psi(t)\rangle$ can be directly evaluated once obtained the joint probability from solving the system \eqref{eq:DiffSys}.
%We also show the analogous case for strongly interacting particles ($U{=}30$) and the intermediate regime ($U{=}1$), respectively in Fig. \ref{fig:1DCarpetsn2}(c) and \ref{fig:1DCarpetsn2}(b). 
We also take into consideration the role of the onsite interaction which affects
the perfect reconstruction of a two particle wavepacket. It turns out that from the space-time dynamics of $\mean{n_j(t)}$ it is not possible to discriminate free evolution ($U{=}0$) from the hard-core limit ($U{=}\infty$), while particle statistics give rise to different dynamics for 
$\mean{ n_j^2(t)}$. 
On the other hand, for intermediate
values of the onsite interaction the dynamics does not lead to a perfect
reconstruction of a wavepacket, due to scattering effects. This effect is clearly shown in Fig. \ref{fig:1DCarpetsn4}(b), \ref{fig:1DCarpetsn4}(e) and in Fig. \ref{fig:Slice2} for $U_n{=}1$ where $\langle n_{2}^2(t{=}2t^*)\rangle {<}\langle n_{2}^2(t{=}0)\rangle$ and $\langle n_{2}(t{=}2t^*)\rangle {<}\langle n_{2}(t{=}0)\rangle$.
%\begin{figure*}
%\caption{Quantum Carpet: the result of four particle interference is shown as $\langle n_j^2(t)\rangle$  in a $L{=}14$ one-dimensional chain for several value of the onsite interaction $U_{n}{=}U$. The initial wavefunction is $\ket{\psi(t{=}0)}{=}a_{2}^{\dagger}a_{6}^{\dagger}a_{9}^{\dagger}a_{13}^{\dagger}\ket{0} $ and $t^*$ is the wavepacket reconstruction time.}
%\label{fig:1DCarpetsn2}
%\end{figure*}
%\begin{figure}[h]
%\includegraphics[width= 8 cm]{n10L40}
%\caption{$\langle n(t)\rangle$ for site $10$ as function of time for a $L{=}40$ chain initialized in $\ket{\psi}{=}a_{10}^{\dagger}a_{31}^{\dagger}\ket{0} $ for several value of the onsite interaction $U$. The curve for two non interacting particles ($U{=}0$)  coincides with the one in the strong inter-particle interaction case ($U{=}30$).}
%\label{fig:MeanNumbSite10}
%\end{figure}
%\begin{figure}[h]
%\includegraphics[width= 8 cm]{nSquare10L40}
%\caption{$\langle n^2(t)\rangle$ for site $10$ as function of time for a $L{=}40$ chain initialized in $\ket{\psi}{=}a_{10}^{\dagger}a_{31}^{\dagger}\ket{0} $ for several value of the onsite interaction $U$.}
%\label{fig:NSquare10}
%\end{figure}
\subsection{Perfect generation of entanglement in an XY spin chain}
We now consider a chain of spin-$\frac12$ magnets described by the XY 
Hamiltonian 
\begin{align}
  %\mathcal H = -\sum_n \left(\frac{J_n}2\sigma_n^+ \sigma_{n+1}^-+{\rm h.c.} \right) -
  \mathcal H = -\sum_n \left({J_n}\sigma_n^+ \sigma_{n+1}^-+{\rm h.c.} \right) -
  \sum_n \frac{B_n}2 \sigma_n^z~,
  \label{e.xx}
\end{align}
where $\sigma^{\alpha}_n$, $\alpha{=}x,y,z$ are the Pauli spin operators acting on
the spin localized in the $n$-th site of the chain, and 
$\sigma^{\pm}_n{=}(\sigma^x_n{\pm} i \sigma^y_n){/}2$. 
Effective spin-$\frac12$ systems coupled by the Hamiltonian \eqref{e.xx} with site dependent
coupling strengths  can be obtained in different physical realizations; 
{\it e.g.} in NRM using global rotations and suitable field gradients 
\cite{ajoy_quantum_2013}, with atomic ions confined in segmented microtraps
\cite{wunderlich_two-dimensional_2009}, 
with neutral atoms trapped into an optical lattice by polarized laser beams 
\cite{duan_controlling_2003,fukuhara_microscopic_2013}, or with superconducting
qubits coupled either by site dependent capacitors \cite{neeley_generation_2010} or inductors \cite{chen_qubit_2014}. 

The system Hamiltonian \eqref{e.xx}
can be mapped to a fermionic hopping model via the Jordan-Wigner
transformation: the operators $c_n{=}\prod_{j<n}(-\sigma^z_j) \sigma_n^-$ 
satisfy canonical anticommutation relations and
$  \mathcal H{=}\sum_{nm} \bra n H\ket m \; c_n^\dagger c_m $~ ,
where $H$ is the hopping matrix \eqref{e.hameng}. 
Every many-body spin state can be obtained by applying the annihilation
operators $c_n$ to the fully polarized state $\ket{\Omega}{=}
\ket{\uparrow\uparrow\cdots}$. Therefore, the time evolution of a 
generic initial state can be obtained by expressing the operator 
$c_n$ in the Heisenberg picture
\cite{banchi_ballistic_2013} as
\begin{align}
c_n(t) = \sum_m \bra n e^{-iHt} \ket m \; c_m. 
  \label{e.cevo}
\end{align}

We now show how one can create entanglement between two remote mirror symmetric
sites by exploiting the perfect wave-packet splitting \eqref{e.transf}.
Suppose that, starting from the fully polarized state $\ket\Omega$ a particle is 
flipped in position $n$; the initial state of the system is then
$c_n\ket\Omega$. When the single-particle Hamiltonian implements the
transformation \eqref{e.transf}, then, thanks to Eq.\eqref{e.cevo} one has 
$c_n\ket\Omega\xrightarrow{t^*} (c_n \ket\Omega {+} ic_{L-n+1}\ket\Omega)/\sqrt
2$. Therefore, going back to the spin picture, after the time $t^*$ an
entangled state $\frac{\ket{\uparrow\downarrow}+i
\ket{\downarrow\uparrow}}{\sqrt 2}$ between sites $n$ and $L{-}n{+}1$
is generated. 

The above arguments can be generalized in a many-particle setting to generate 
the maximal amount of entangled pairs starting from a separable state.  
Two suitable choices of the initial state are 
\begin{align}
  \ket{\psi^{\rm DM}}&=\ket{\uparrow\uparrow\cdots\uparrow\downarrow\cdots\downarrow
  \downarrow}~,
  \\
  \ket{\psi^{\rm AFM}} &=
  \ket{\uparrow\downarrow\uparrow\cdots\uparrow\downarrow}~,
\end{align}
namely the domain-wall state 
$ \ket{\psi^{\rm DM}}$ or the anti-ferromagnetic state $\ket{\psi^{\rm AFM}}$.
If the system is initialized in either $ \ket{\psi^{\rm DM}}$  or
$\ket{\psi^{\rm AFM}}$ 
and is let to evolve under the perfect splitting Hamiltonian, then
the resulting state after a time $t^*$ 
is $
(c_1+ e^{i\alpha_1}c_L)
(c_2+ e^{i\alpha_2}c_{L-1})
(c_3+ e^{i\alpha_2}c_{L-2})\cdots\ket\Omega$, where $\alpha_i$ depends on the 
initial state. By carefully dealing with the Jordan-Wigner phase entering into
the definition of the operators $c_n$ one can easily find that the resulting
state corresponds to a state in which every pair of qubits sitting in positions
$n$ and $L{-}n{+}1$ is maximally entangled. 
The perfect splitting dynamics thus represents an alternative to other methods
existing in the literature to generate 
nested Bell pairs \cite{di_franco_nested_2008,alkurtass_optimal_2014} starting from a separable state. However, compared to previous proposals it is more general because it allows tuning the number of generated Bell pairs by simply choosing the number of flipped
spins in the initial state. 

\section{Conclusions}
In this paper we study the wavefunction dynamics of hopping particles and/or
quasi-particles in a quantum chain. We design the Hamiltonian so that 
a localized wave packet evolves coherently along the chain without dispersion, 
and at particular
point is perfectly split into transmitted and reflected components
which propagate in opposite directions without dispersion. 
When the reflected component reaches the initial site, its wave packet becomes
localized  while, 
at the same time, the wave packet of the transmitted component
becomes localized in a different site of the chain. 
We devise the exact conditions that the Hamiltonian spectrum has to satisfy to
allow for the perfect splitting and reconstruction. 
Then we focus on some viable Hamiltonians with
nearest-neighbor interactions and site-dependent couplings, and we find the 
coupling pattern which satisfies the perfect splitting condition using 
inverse eigenvalue techniques. 

Besides shedding new light into quantum interference phenomena in one
dimension, our results are particularly useful for applications. 
In this respect, we study atomic lattices and obtain perfect Hanbury Brown and
Twiss correlations and peculiar quantum interference patterns which result in 
regular structure in the space-time evolution of the many-particle wave
function. 
Moreover, we show that in a spin chain setting, the particle splitting can be
used to generate maximally entangled states between distant parts. 

We expect that the perfect wavepacket splitting will become a general tool for 
varied applications in controlled quantum interference and quantum information
processing.

\section{Acknowledgments}
The authors %are currently supported by the ERC grant PACOMANEDIA. 
acknowledge the financial support by the ERC under Starting Grant 308253
PACOMANEDIA. 

\appendix

\section{Perfect splitting Hamiltonian for $L=5,6$}
The Hamiltonian matrices $\bra n H \ket m$ for perfect balanced splitting when 
$L{=}5$ and $6$ are respectively (in unit of $J$):
\begin{align*}
  \left(
  \begin{array}{ccccc}
     -0.08378 & 0.6195 & 0 & 0 & 0 \\
      0.6195 & -0.2932 & 0.6664 & 0 & 0 \\
       0 & 0.6664 & 0.7540 & 0.6664 & 0 \\
        0 & 0 & 0.6664 & -0.2932 & 0.6195 \\
         0 & 0 & 0 & 0.6195 & -0.08378 \\
       \end{array}
       \right)
       \\
       \left(
       \begin{array}{cccccc}
          0 & 0.5999 & 0 & 0 & 0 & 0 \\
           0.5999 & 0 & 0.8279 & 0 & 0 & 0 \\
            0 & 0.8279 & 0 & 0.3927 & 0 & 0 \\
             0 & 0 & 0.3927 & 0 & 0.8279 & 0 \\
              0 & 0 & 0 & 0.8279 & 0 & 0.5999 \\
               0 & 0 & 0 & 0 & 0.5999 & 0 \\
             \end{array}
             \right)~.
\end{align*}
As shown in Section III, small imperfections parameter tuning result in negligible deviations from the ideal dynamics.

\end{document}